\documentclass[11pt,twocolumn,twoside]{article}
\usepackage{fully3d}
\usepackage{algorithm}
\usepackage{algpseudocode}
\usepackage{subcaption}

\graphicspath{{images_final}}

\usepackage{amssymb}
\usepackage{multirow}

\DeclareMathOperator*{\Id}{Id}
\DeclareMathOperator*{\Prox}{Prox}

\DeclareMathOperator*{\argmin}{arg\,min}
\newcommand{\xb}{\ensuremath{\mathbf{x}}}
\newcommand{\yb}{\ensuremath{\mathbf{y}}}
\newcommand{\ub}{\ensuremath{\mathbf{u}}}
\newcommand{\zb}{\ensuremath{\mathbf{z}}}
\newcommand{\thetab}{\ensuremath{\boldsymbol{\theta}}}

\begin{document}

\title{Convergent ADMM Plug and Play PET Image Reconstruction} 

\author[1]{Florent~Sureau}
\author[1,2]{Mahdi~Latreche}
\author[1]{Marion~Savanier}
\author[1]{Claude~Comtat}

\affil[1]{BioMaps, Université Paris-Saclay, CEA, CNRS, Inserm, SHFJ, 91401 Orsay, France.}

\affil[2]{Institut Denis Poisson, Université d’Orléans, UMR CNRS 7013, 45067 Orléans, France.}

\maketitle
\thispagestyle{fancy}


\begin{customabstract}
In this work, we investigate hybrid PET reconstruction algorithms based on coupling a model-based variational reconstruction and the application of a separately learnt Deep Neural Network operator (DNN) in an ADMM Plug and Play framework. Following recent results in optimization, fixed point convergence of the scheme can be achieved by enforcing an additional constraint on network parameters during learning. We propose such an ADMM algorithm and show in a realistic $[{}^{18}\text{F}]$-FDG synthetic brain exam that the proposed scheme indeed lead experimentally to convergence to a meaningful fixed point. When the proposed constraint is not enforced during learning of the DNN, the proposed ADMM algorithm was observed experimentally not to converge.
\end{customabstract}


\section{Introduction}

In the quest for PET image reconstructions adapted to a protocol, to a specific patient or to a given task, deep learning approaches are currently a promising avenue of research. Early works have already illustrated the potential of such approaches to achieve better resolution, contrast recovery and noise propagation trade-offs compared to classical model-based variational methods for PET reconstruction \cite{Gong2019c,Mehranian2021}. In particular, a reduction of the dose injected to the patient could be envisioned without sacrificing much of the reconstructed image quality compared to a standard dose exam, which would be beneficial for the patient and/or to reduce the cost of a PET scan \cite{Kaplan2018}. 

However using deep learning in PET raises several new issues compared to the aforementioned variational approaches. In particular these specific methods can generate instabilities related to the ill-posedness of the problem that could lead to images with artefacts as already observed in biomedical image reconstruction applications \cite{Antun2020}. Furthermore, deep learning techniques for reconstruction often use neural networks as a black box operator in the reconstruction pipeline. This leads to end estimates that lack mathematical or statistical guarantees that could make them robust, contrary to classical reconstruction techniques typically associated with a convex variational problem. Robustness of the reconstruction is also of paramount importance in PET where datasets are often small (with typically only tens of exams per protocol) leading to limited learning and validation.

In this context, hybrid techniques inspired by model-based PET reconstruction approaches in a learning framework have been proposed to avoid learning the direct model and to aim at more reliable estimates \cite{Reader2021}. 
 We will focus on this work on a less investigated hybrid framework for PET reconstruction than unrolled \cite{Mehranian2021} or synthesis \cite{Gong2019c} approaches: the ADMM Plug and Play approach \cite{Venkatakrishnan2013}. In this framework, an implicit operator related to the prior is only learned \cite{Meinhardt2017}, making this approach flexible. Fixed-point convergence results have been investigated for this framework even though the optimization problem is non-convex and the learned operator is implicit \cite{Chan2017,Ryu2019}. Based on these results, we propose and investigate in this work a convergent ADMM Plug and Play approach for PET image reconstruction using Deep Learning. We present in section~\ref{sec:materials} the convergent ADMM Plug and Play approach that we propose, the datasets used for training and evaluation and detail the architecture and implementation of the additional constraint needed for fixed-point convergence of the scheme. We then present and discuss our results on realistic $[{}^{18}\text{F}]$-FDG synthetic exams. 

\section{\label{sec:materials}Materials and Methods}

\subsection{Convergent ADMM Plug and Play}
We consider the reconstruction of an image denoted by $\xb \in\mathbb{R}_+^N$, from an observed noisy sinogram $\yb\in\mathbb{N}^M$. The ADMM plug and play algorithm is described in Algorithm~\ref{algo:ADMM_PnP}. In our context of PET reconstruction $LL(\yb,\xb)$ is the Poisson log-likelihood and  $\textcolor{red}{\mathcal{D}_{\thetab} }$ is a DNN operator with inputs the reconstructed PET images, with parameters learnt in a separated step.

\begin{algorithm}
\caption{ADMM Plug and Play with a DNN $\textcolor{red}{\mathcal{D}_{\thetab} }$.\label{algo:ADMM_PnP}}
\begin{algorithmic}[1]
	\State Choose $\zb^{(0)}$, $\ub^{(0)}$, $\textcolor{green}{\boldsymbol{\rho}}$, $K$.
	\For{$k=0..K$}
	\State	$\xb^{(k+1)}_\rho=\argmin\limits_{\xb\in\mathbb{R}_+^N} -LL(\yb,\xb) +\frac{\textcolor{green}{\boldsymbol{\rho}}}{2}\|\xb-(\zb^{(k)}_\rho-\ub^{(k)}_\rho)\|^2_2$  \label{eq:recons} 
	\State $\zb^{(k+1)}_\rho={\textcolor{red}{\mathcal{D}_{\thetab }}}\left(\xb^{(k+1)}_\rho+\ub^{(k)}_\rho\right)$
		\State $\ub^{(k+1)}_\rho=\ub^{(k)}_\rho+\xb^{(k+1)}_\rho-\zb^{(k+1)}_\rho$
	\EndFor
	\State \Return $\xb^{(K+1)}_\rho$
\end{algorithmic}
\end{algorithm}

Compared to unrolling techniques, this ADMM framework decouples learning the DNN and reconstructing the images, which is convenient for integrating DNN inside the reconstruction. 
In particular the number of ADMM iterations can be large with no impact on the GPU VRAM contrary to unrolling techniques.
Besides the reconstruction problem in line~\ref{eq:recons} is a standard (convex) minimization problem, which can be solved using an efficient existing PET algorithm \cite{DePierro1995a}. Compared to synthesis ADMM approaches, this Plug and Play formulation leads to a simple condition on the DNN for this scheme to converge as described below. Note in particular that no heuristic choice of an iteration-dependent $\rho$ is needed to stabilize the algorithm as often proposed as in \cite{Chan2017}, even though our final estimate still depends on $\rho$.

If the operator $\textcolor{red}{\mathcal{D}_{\thetab}}$ corresponds to the proximal operator of a proper, closed and convex function, Algorithm~\ref{algo:ADMM_PnP} is the classical ADMM algorithm with an implicit operator and with convergence properties described in \cite{Eckstein1992,Boyd2010} . However for a general $\textcolor{red}{\mathcal{D}_{\thetab}}$, such algorithm may not even minimize a convex problem and there is no guarantee of convergence of such a scheme. Conditions for fixed point convergence of the ADMM Plug and Play algorithm have however been studied in \cite{Chan2017,Ryu2019} and references therein. In particular, \cite{Ryu2019} shows that one ADMM iteration can be written equivalently as applying an operator $\mathcal{T}_{\thetab }=\frac{1}{2} \Id + \frac{1}{2} (2{\textcolor{red}{\mathcal{D}_{\thetab }}}-\Id)(2 \Prox_{-LL/\rho}-\Id)$. This implies in particular that if  $\mathcal{L}_{\thetab }=(2{\textcolor{red}{\mathcal{D}_{\thetab }}}-\Id)$ is a non expansive operator and  $\mathcal{T}_{\thetab}$ has a fixed point, then fixed point convergence of the scheme is obtained \cite{Ryu2019}. The non-expansiveness constraint is however difficult to enforce numerically. \cite{Ryu2019} proposed to use real spectral normalization on each layer 
to constrain the Lispchitz constant of each layer of the DNN. In this work, we rather use the approach proposed in \cite{Pesquet2021}.
Taking also into account a supervised loss (Mean Squared Error - MSE), the DNN parameters are estimated in the following minimization problem:
\begin{equation}
	\min\limits_{\thetab} \underbrace{\sum\limits_{b=1}^B \|{\textcolor{red}{\mathcal{D}_{\thetab}}}(\mathbf{x}_b)-\bar{\mathbf{x}}_b \|^2}_{\text{MSE}}	+\beta\underbrace{\max\{\|\nabla\mathcal{L}_{\thetab}(\tilde{\mathbf{x}}_b)\|+\epsilon-1,0\}^{1+\alpha}}_{\text{Non expansiveness constraint}},\label{eq:loss}	
\end{equation}
where $b$ is the batch index, $\epsilon$, $\alpha$ and $\beta$ are hyperparameters to balance the supervised loss and the constraint and $\tilde{\xb}_b$ is obtained as a random convex combination of the reference image $\bar{\xb}_b$ and the output of the neural network as follows:
\begin{equation}
	\tilde{\xb}_b=\kappa\bar{\xb}_b+(1-\kappa){\textcolor{red}{\mathcal{D}_{\thetab}}}(\xb_b), \kappa\sim\mathcal{U}[0,1]. \label{eq:sampling}
\end{equation}
Note that the spectral norm of the Jacobian for a given entry point can be estimated using automatic differentiation but the computation is particularly intensive in terms of both GPU VRAM and execution time.

\subsection{Datasets and Learning settings}

The database used for learning and evaluation of the proposed approach was derived from 14 brain $[{}^{18}\text{F}]$-FDG brain exams of healthy subjects and their associated T1 weighted MR images. 
The T1 images were first segmented into 100 regions using FreeSurfer\footnote{\url{https://surfer.nmr.mgh.harvard.edu}}. 
The PET signal was then measured in a frame between 30 minutes and 60 minutes after injection in each region using PETSurfer \cite{Greve2016} to generate 14 distinct anatomo-functional phantoms. 3-dimensional PET simulations for a Biograph 6 TruePoint TrueV PET system were then generated using an analytical simulator \cite{Stute2015}, including normalization, attenuation, scatter and random effects. 
11 phantoms were used for training and 3 for testing. 
Data augmentation was performed for each phantom by simulating 10 realizations of the injected dose so that the total number of counts simulated spans the range observed in the 14 exams. 
This results in 110 shuffled simulations used for training and 30 realizations for testing. 
These simulations were reconstructed with CASToR \cite{Merlin2018} using OSEM with 8 iterations of 14 subsets.

The network $\textcolor{red}{\mathcal{D}_{\thetab}}$ was chosen as a U-Net \cite{Ronneberger2015} with 443649 parameters.
We made several modifications compared to the architecture presented in \cite{Gong2019c}, as a balance between performance of the network and number of parameters to learn: we use 3 levels with instance normalization, 3D average pooling and concatenation between the decoder branches and encoder branches, and we use an overall skip connection to learn on the residual image. Note that the input reconstructed images are first normalized so that the network performance is robust to dose variation. The normalization factor is then applied to the output of the network to recover the correct scale.

The DNN parameters of the U-Net were learned in two steps. In a preliminary phase, the DNN parameters are learnt only with the supervised MSE loss. 
The ADAM optimizer with 50 epochs and a learning rate of 0.001 was used. Batch size was 1, and the reference in the supervised loss corresponds to the  noise-free images.  This results on a first DNN without the constraint on the Jacobian, named "PRE" in the following. 
In a second phase, the total loss in Equation~\ref{eq:loss} is considered and the Power Iterative Method (with a maximum of 10 iterations) and automatic differentiation is used to compute the spectral norm of the Jacobian. In this case we use 14 additional epochs on the PRE DNN to enforce the constraint, using ADAM with a learning rate of 0.0005, batch size of 5, $\beta=10$, $\alpha=0.1$ and  $\epsilon=0.05$ in Equation~\ref{eq:loss}. This network is named "JAC" in the following. 

Both networks are then employed in Algorithm~\ref{algo:ADMM_PnP} using 40 iterations and compared on the simulations. For initialization $\zb^{(0)}$ is an OSEM reconstructed image with 8 iterations of 14 subsets and $\ub^{(0)}=\mathbf{0}$. We first investigated the choice of $\rho$ on the convergence speed and on the solution by looking at the norm of the primal residual defined as $\xb^{(k)}_\rho-\zb^{(k)}_\rho$ and of the dual residual $\rho(\zb^{(k+1)}_\rho-\zb^{(k)}_\rho)$ \cite{Boyd2010}. Both should converge to zero for ADMM to converge.

\section{Results}

\begin{figure}	
	\centering
	\includegraphics[width=0.3\linewidth]{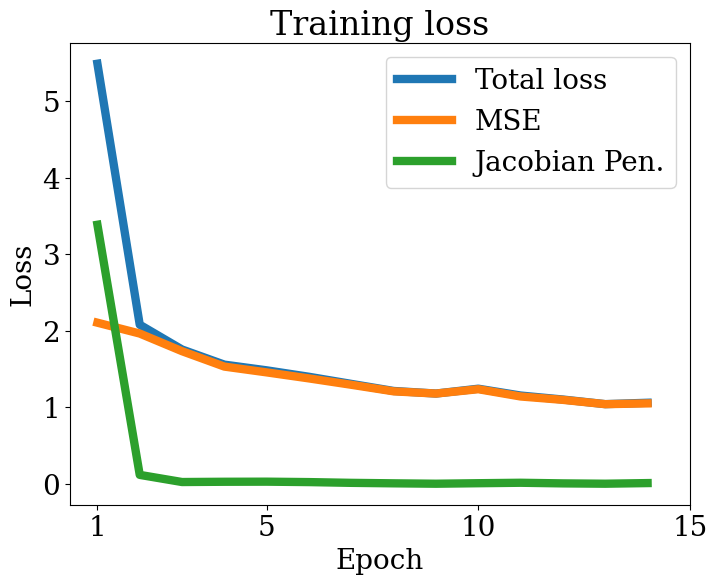}
	\includegraphics[width=0.3\linewidth]{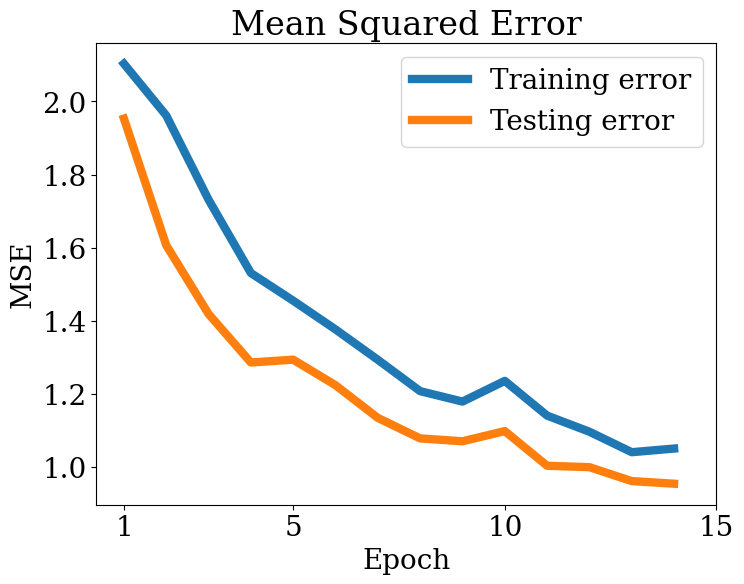}
	\includegraphics[width=0.3\linewidth]{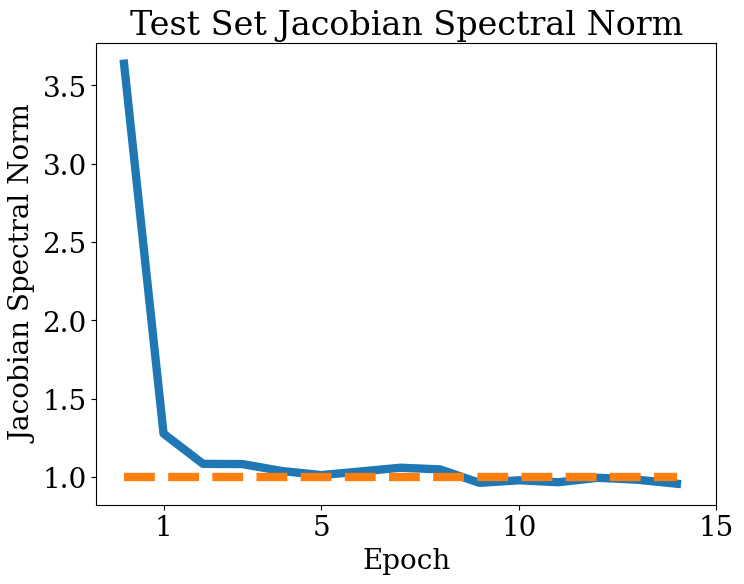}
	\caption{Loss functions for JAC U-Net. Left: the total loss is subdivided into its two individal contribution. Middle: MSE for training and testing phases. Right: Jacobian spectral norm for testing dataset.}
	\label{fig:JAC_results}
\end{figure}

MSE curves during training and testing of the PRE U-Net illustrate that 30 epochs are necessary to learn parameters in the preliminary phase (not shown).  Figure~\ref{fig:JAC_results} shows that the choice of $\beta$ leads to balanced supervised and constraint loss in the first iteration. Both MSE and Jacobian constraint components of the loss are decreasing over epochs. Performance in the testing and training datasets were comparable in terms of MSE, and the proposed implementation of the constraint leads to a Jacobian spectral norm in the testing dataset less than 1 as expected (compared to more than 3.5 for PRE). 

\begin{figure}	
	\centering
	\includegraphics[width=0.75\linewidth]{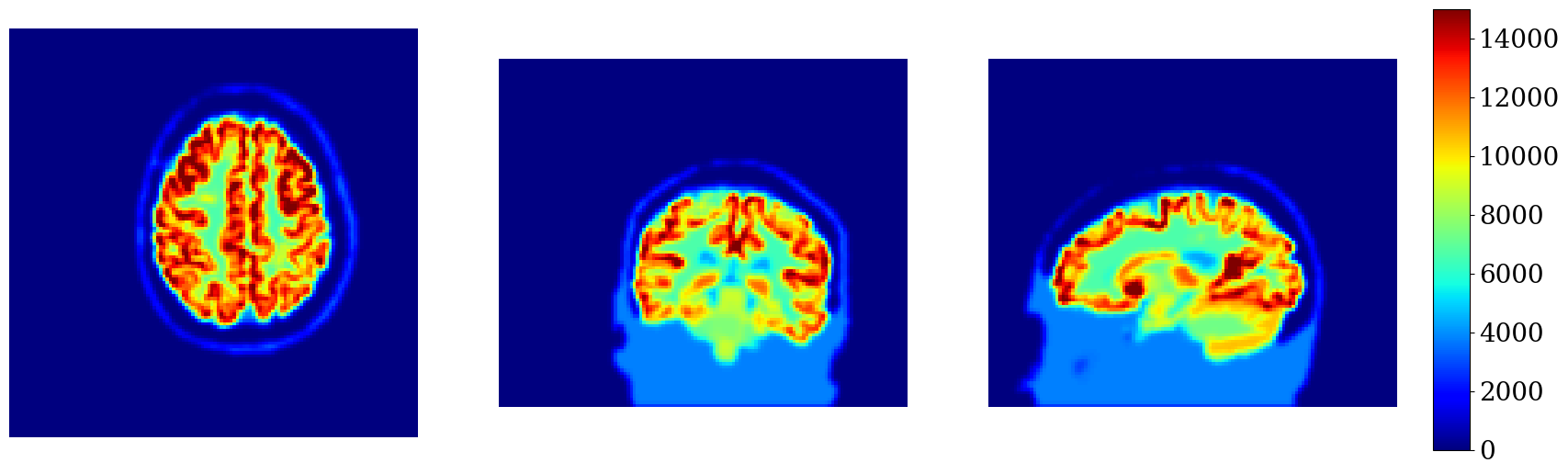}
	\includegraphics[width=0.75\linewidth]{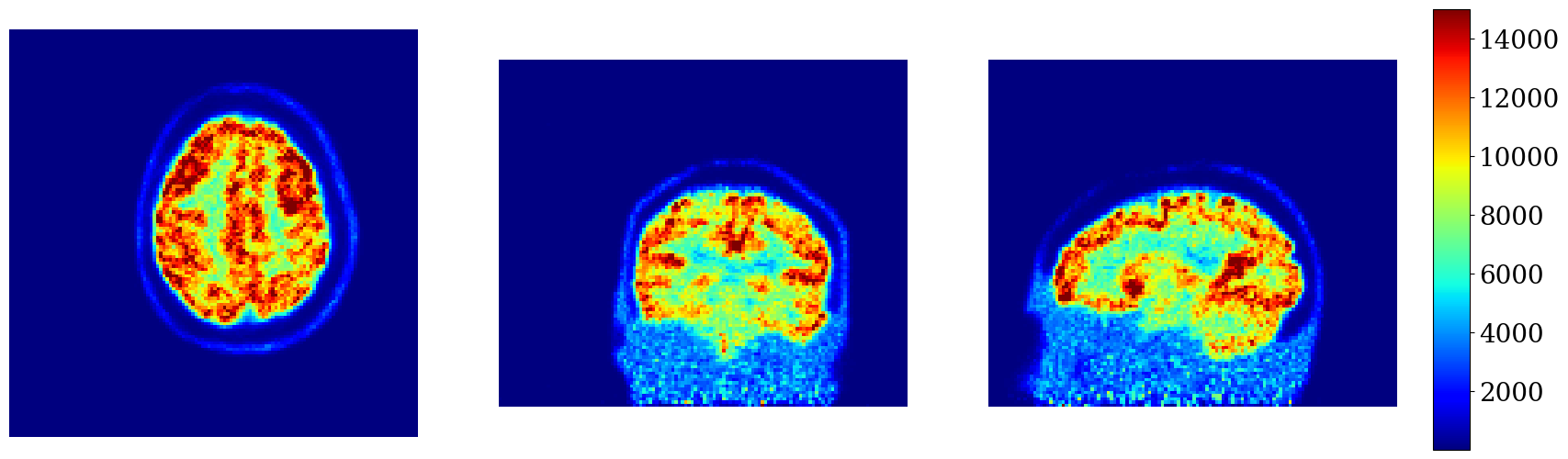}
	\includegraphics[width=0.75\linewidth]{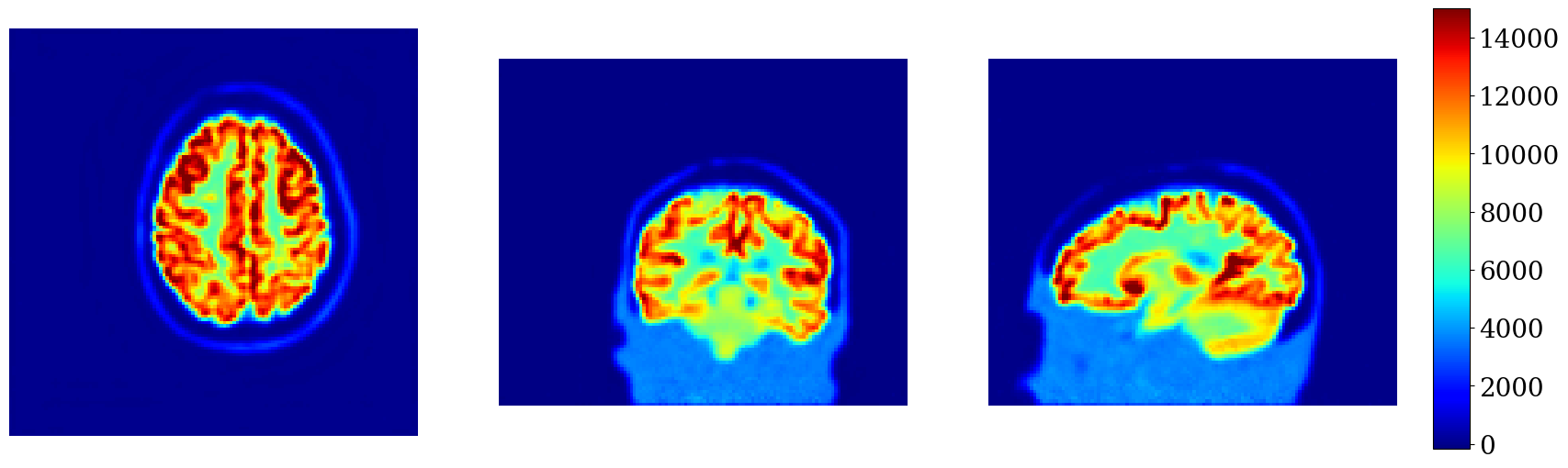}
	\includegraphics[width=0.75\linewidth]{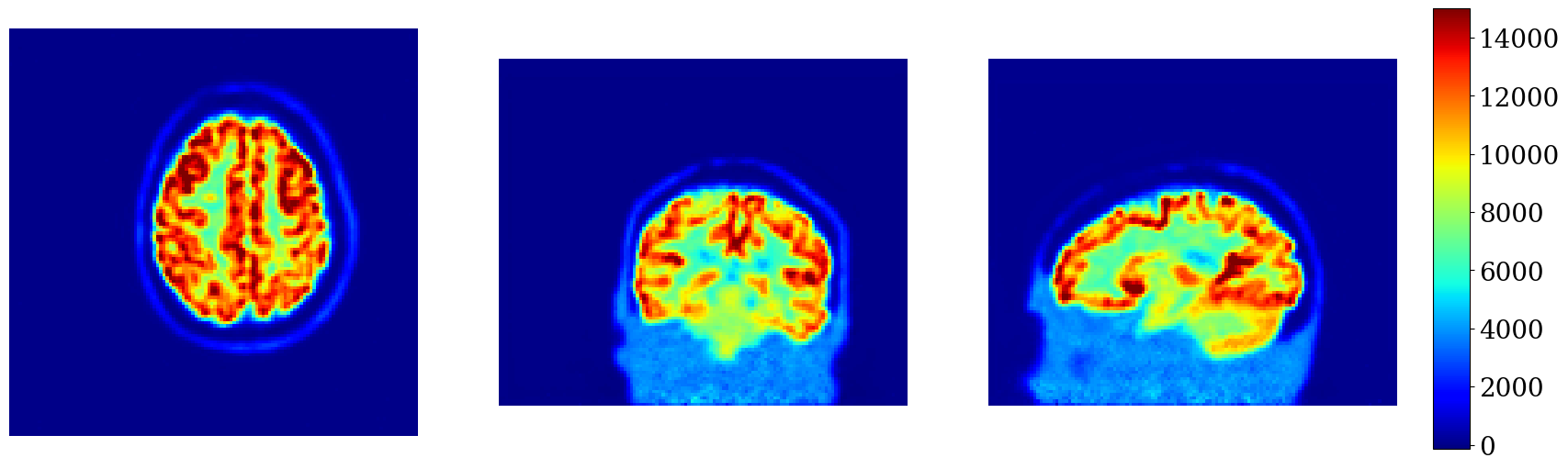}
	\caption{Performance of U-Nets in post-processing for a simulation in the test set. From top to bottom: noise-free reference image, OSEM image, PRE U-Net estimate, JAC U-Net estimate.}
	\label{fig:PREJAC_results}
\end{figure}

Figure~\ref{fig:PREJAC_results} illustrates the performance of both U-Nets when used as simple post-processing for reconstructed PET images: propagated noise has been reduced while preserving the high frequency structures present in the original phantom. Compared to the PRE U-Net higher frequencies are observed in the background region for the JAC U-Net, indicating that the denoising performance is slightly degraded when using the Jacobian constraint.  

\begin{figure}	
	\centering
	\includegraphics[width=0.4\linewidth]{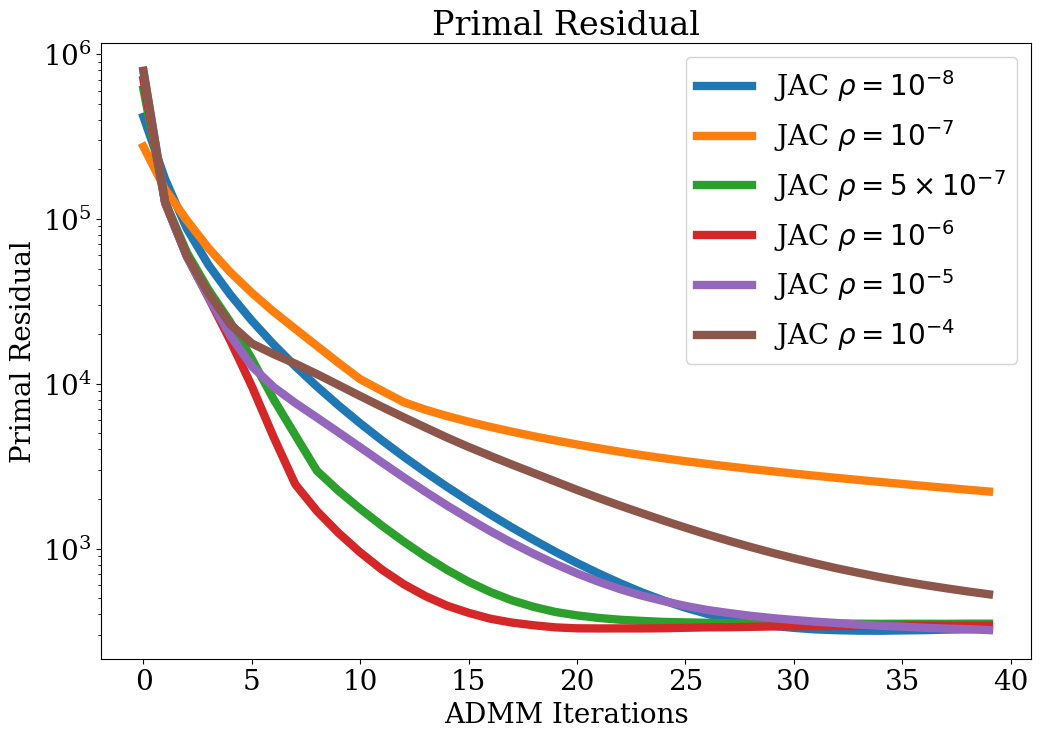}
	\includegraphics[width=0.4\linewidth]{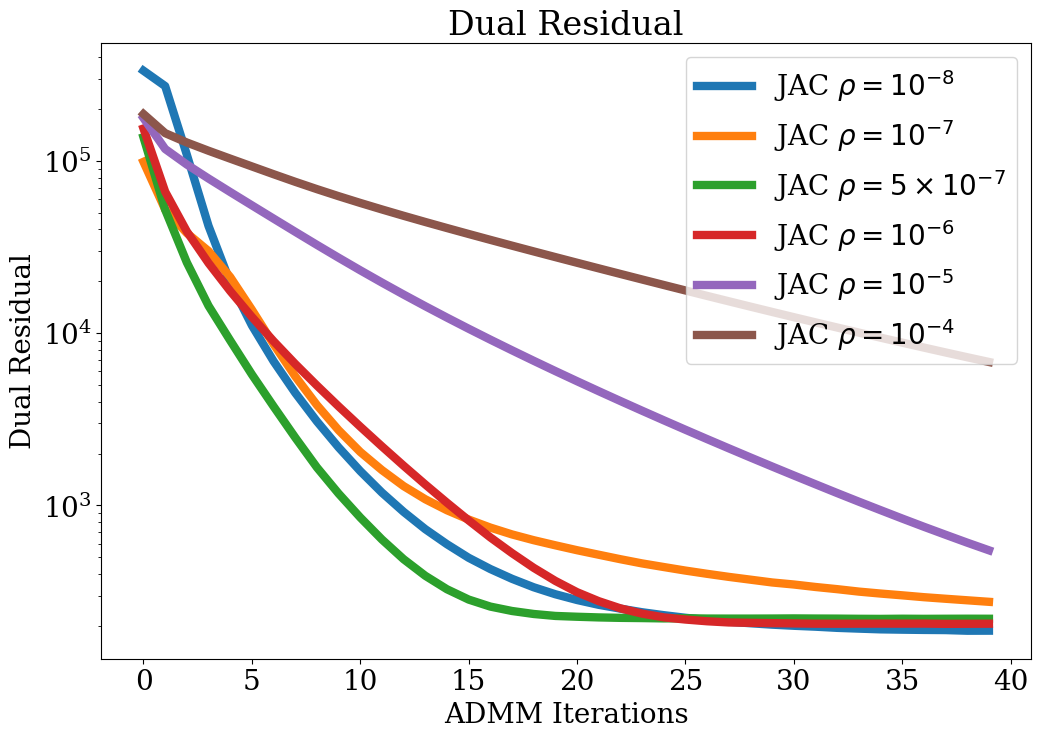}
	\caption{Norm of the primal (left) and dual residual (right) across ADMM Plug and Play iterations for the JAC U-Net.}
	\label{fig:ADMM_rho}
\end{figure}

The two networks were then employed in Algorithm~\ref{algo:ADMM_PnP}. The impact of hyperparameter $\rho$ is illustrated on Figure~\ref{fig:ADMM_rho} for the JAC U-Net. This illustrates that a carefull choice of this hyperparameter is needed to reach adequate convergence speed for the overall scheme similarly to what is observed in ADMM in convex problems. In the following, we chose $\rho=5e-7$ which achieves fast convergence as indicated in both primal and dual residuals.

\begin{figure}	
	\centering
	\begin{tabular}{cc}
	\includegraphics[width=0.35\linewidth]{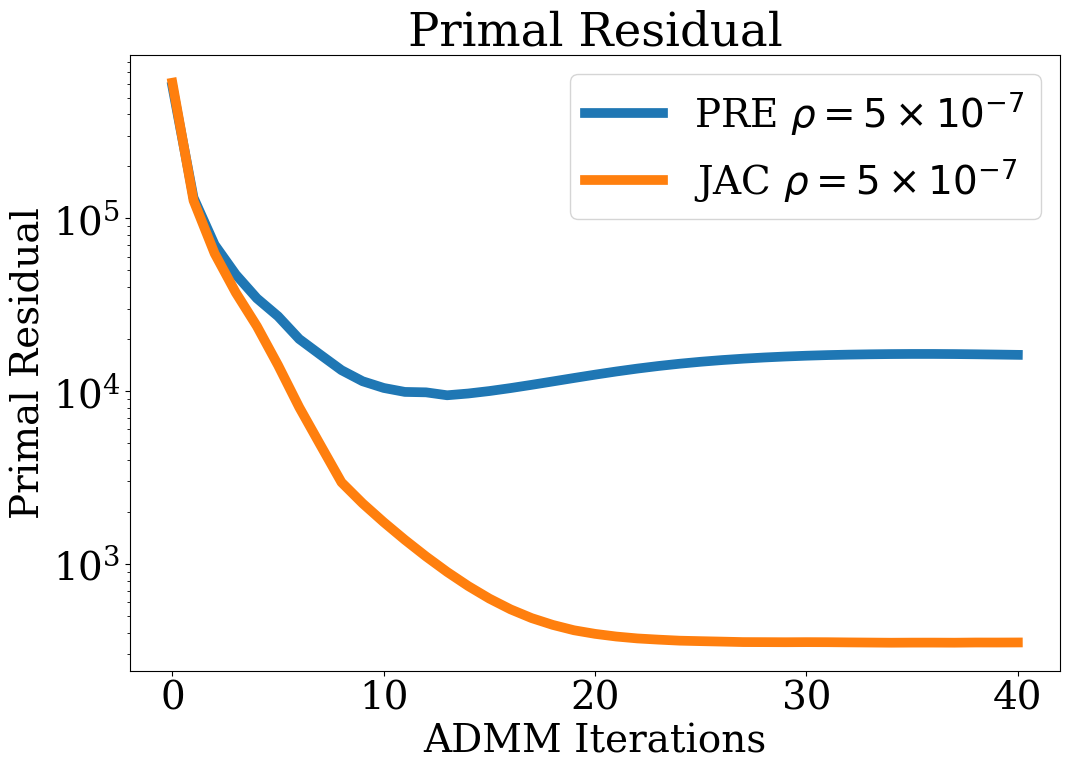}&
	\includegraphics[width=0.35\linewidth]{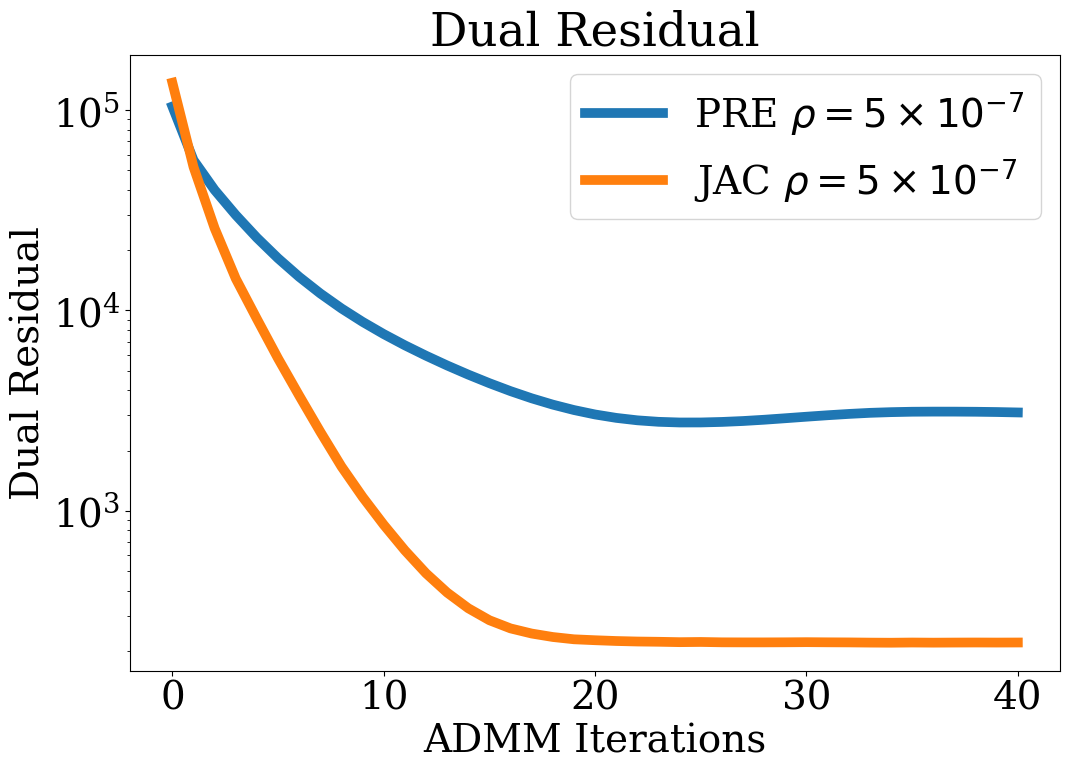}\\
	\includegraphics[width=0.4\linewidth]{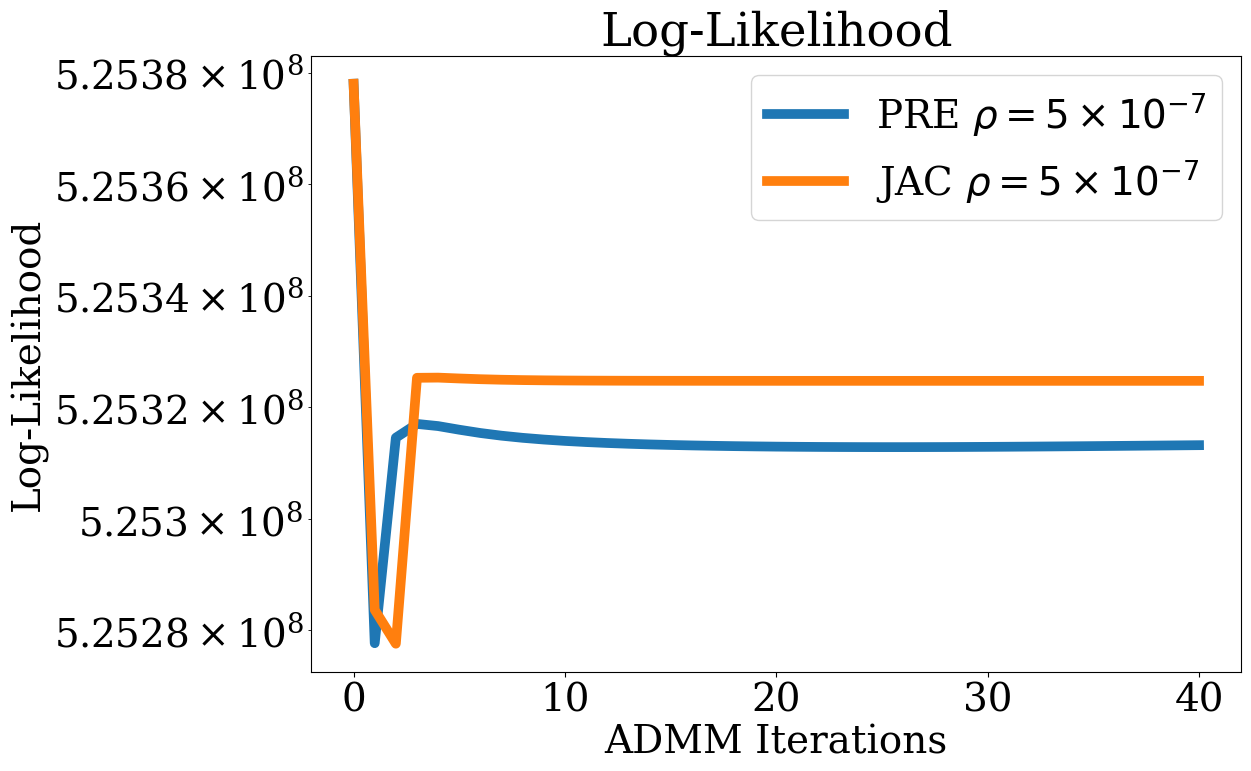}&
	\includegraphics[width=0.35\linewidth]{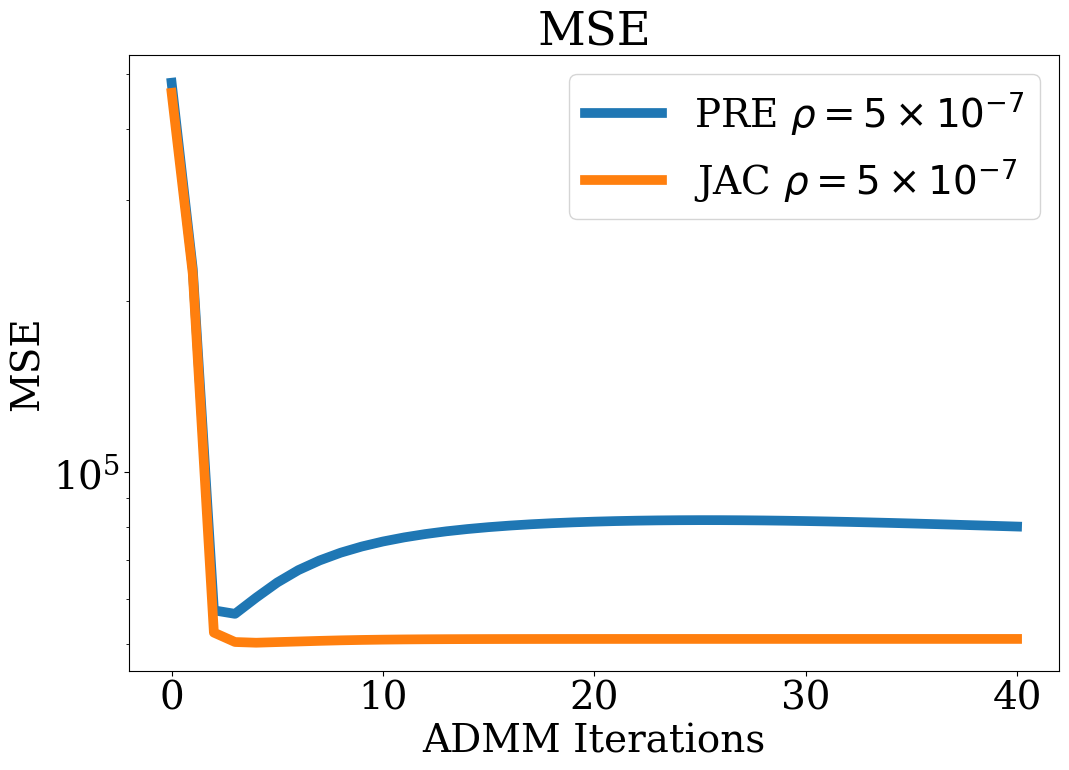}
	\end{tabular}
	\caption{Evolution for both U-Nets of Top: primal (left) and dual (right) residuals; Bottom: log-likelihood (left) and MSE (right).}
	\label{fig:ADMM_JACREG_LOSS}
\end{figure}

Figure~\ref{fig:ADMM_JACREG_LOSS} illustrates the performance of PRE and JAC U-Nets across ADMM iterations, for the previously selected value of $\rho$. It can be observed that JAC U-Net leads to decreasing primal and dual residuals and to a rapid stabilization of both MSE to a low value and log-likelihood to a high value. On the contrary the PRE U-Net has not converged as illustrated in primal and dual residuals, and has lower log-likelihood and higher MSE than JAC. 

\begin{figure}	
	\centering
	\includegraphics[width=0.8\linewidth]{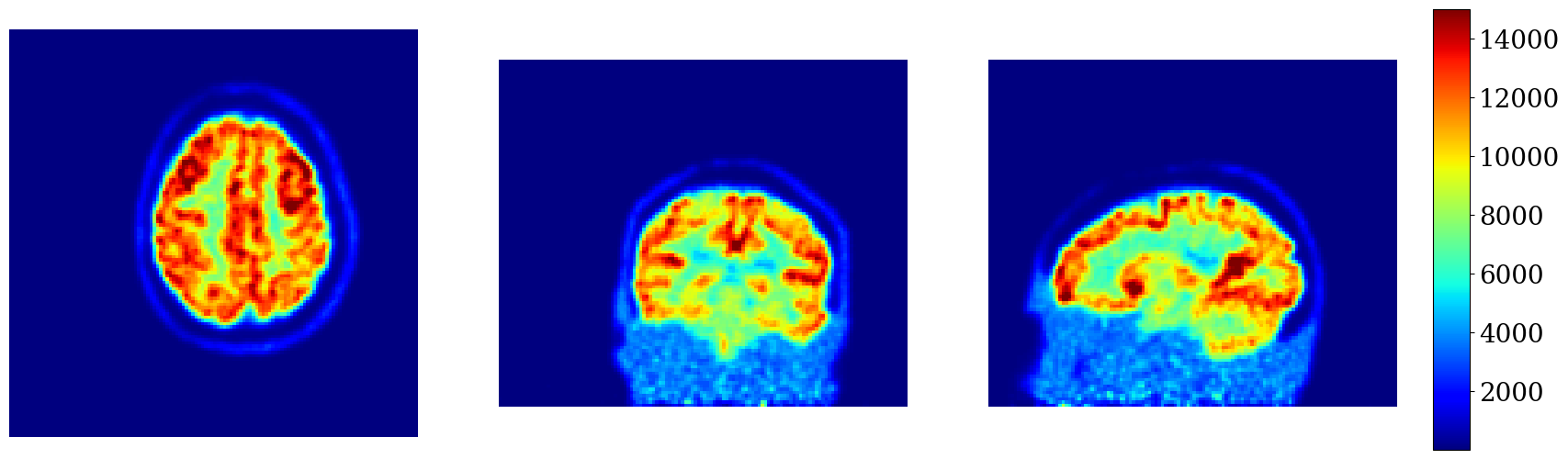}
	\includegraphics[width=0.8\linewidth]{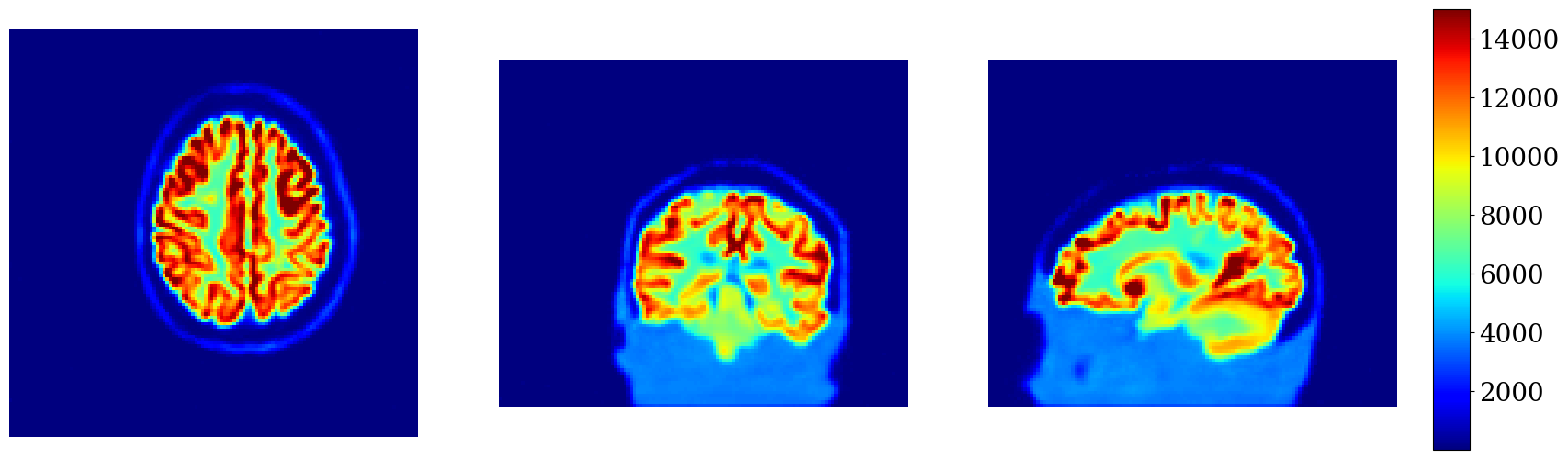}
	\includegraphics[width=0.8\linewidth]{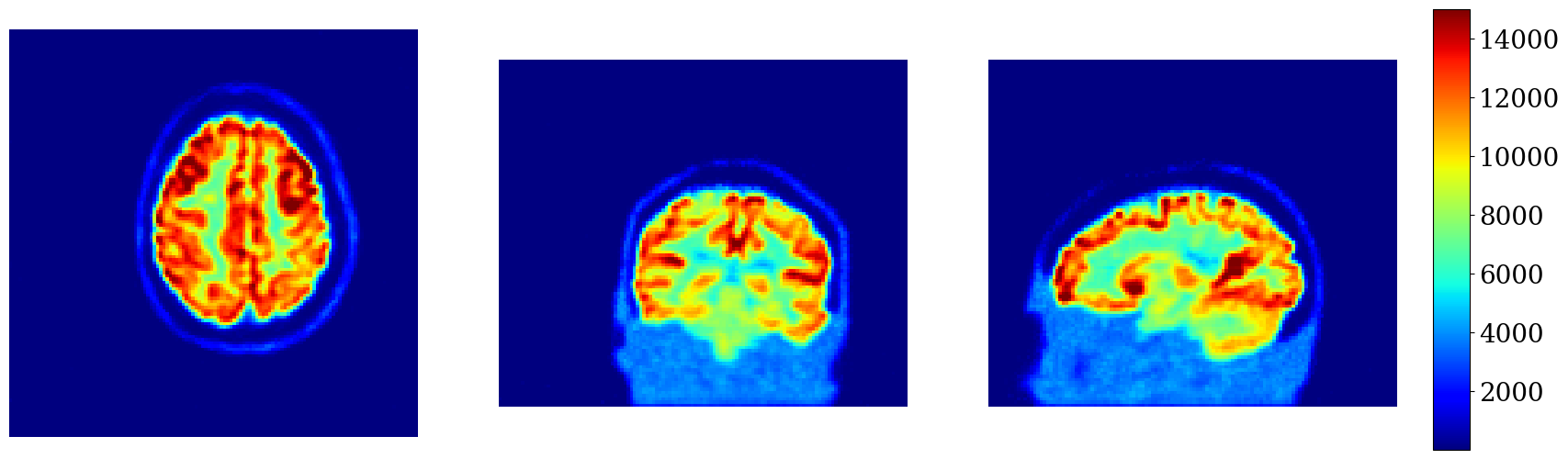}
	\caption{Reconstructed images for the same test set simulation as in Figure~\ref{fig:PREJAC_results}. On top, the OSEM reconstructed image with the best Gaussian post-filtering selected (MSE:90265). In the middle, the PRE U-Net results after 80 iterations (MSE:75227). In the last row, the result obtained with the proposed JAC U-Net after 80 iterations (MSE:50957).}
	\label{fig:ADMM_JACREG_IM}
\end{figure}

Figure~\ref{fig:ADMM_JACREG_IM} illustrates the recovered images using ADMM Plug and Play with PRE and JAC U-Nets, compared to the best Gaussian post-filtered OSEM image. JAC results lead to the closest image to the reference, with the lowest MSE. On the contrary, PRE U-Net leads to a not converged image further from the reference. 

\section{Discussion}

In this work we have proposed a strategy to build a convergent ADMM Plug and Play algorithm by enforcing a non-expansiveness constraint during the learning of the DNN. Enforcing a strict (global) non-expansiveness constraint is actually replaced by enforcing $\|\nabla\mathcal{L}_{\thetab}(\tilde{\mathbf{x}}_b)\|$ to be less than 1 for sampled $\tilde{\mathbf{x}}_b$ using Equation~\ref{eq:loss}. More epochs are needed to ensure sufficient sampling of the space close to the solution. However, the results presented in this work illustrate that the constraint is satisfied even for the test set. It was also observed experimentally that Algorithm~\ref{algo:ADMM_PnP} converges as expected. Nonetheless, robustness of such a reconstruction scheme should be further assessed.
We have shown that the choice of hyperparameter $\rho$ is crucial for convergence speed as in the convex case, but in this non-convex case the solution also depends on $\rho$. The choice of $\rho$ is therefore crucial, and the choice of this hyperparameter should be investigated more thoroughly. Finally we plan to investigate the performance of such algorithm in low-dose scenarios to assess the performance of such approach in a more clinically relevant setting.

\section{Conclusion}

In this work we have proposed a new approach for PET reconstruction using Deep Learning. Based on the ADMM Plug and Play framework, the proposed approach uses a constraint on the spectral norm of an operator Jacobian during learning. This promotes the non-expansiveness that leads to a convergent reconstruction scheme. We show in experimental simulations that without this constraint the ADMM does not converge. On the contrary, the proposed approach experimentally converges to a higher likelihood solution and to a lower MSE, illustrating the interest of such an approach.

\section{Acknowledgment}

We acknowledge financial support from the French National Research Agency (ANR) under grant ANR-20-CE45-0020 (ANR MULTIRECON). This work was partly funded by the France Life Imaging (ANR-11-INBS-0006 grant from the French “Investissements d’Avenir” program).

{\footnotesize\printbibliography}

\end{document}